\documentstyle[myagu,epsfig]{article}

\lefthead{TSERKOVNYAK JOHNSON}

\righthead{CAN ONE HEAR THE SHAPE OF A SATURATION PATCH?}

\received{}
\revised{}
\accepted{}

\paperid{}
\cpright{}{}
\ccc{}

\authoraddr{Y. I. Tserkovnyak,
Lyman Laboratory of Physics, Harvard University, Cambridge, MA 02138
(e-mail: yaroslav@physics.harvard.edu)}
\authoraddr{D. L. Johnson, Schlumberger-Doll Research, Old Quarry Road,
Ridgefield, CT 06877-4108 (e-mail: thesound@ridgefield.sdr.slb.com)}

\begin{document}
\today
\title{Can one hear the shape of a saturation patch?}
\author{Yaroslav Tserkovnyak}
\affil{Harvard University, Lyman Laboratory of Physics, Massachusetts}
\author{David Linton Johnson}
\affil{Schlumberger-Doll Research, Connecticut}

\begin{abstract}
The theory of the acoustics of patchy-saturation in porous media
is used to analyze experimental data on wave velocity and
attenuation in partially water saturated limestones.  It is
demonstrated that the theory can be used to deduce the value of
$V/A$, the ratio of the volume to area of the water patch, and
$l_f$, the Poisson size of the water patch.  One can ``hear'' the
shape of a patch if the properties of the rock and the measurement
frequencies are such as to satisfy the specific requirements for
the validity of the theory.
\end{abstract}

\begin{article}

The Biot theory \cite{biot54} can be used to describe the acoustic
properties of an elastic porous solid fully-saturated with a
compressible Newtonian fluid. (For a review see \cite{johnson86}
and references therein.)  The input parameters of the theory are
directly measurable on a given sample by independent means.  The
success of the theory in predicting the dispersion and the
attenuation of the fast compressional, slow compressional, and
shear modes has been demonstrated in a number of relevant
experiments (\cite{johnson94} and references therein).

\markcite{{\it Johnson}} [2001] has extended this theory to the
case of an elastic porous frame containing two different fluids
filling the entire frame in non-mixing ``patches" (i.e. at every
point of the sample, the frame is fully-saturated with one of the
two fluids).  The theory is a simplification and generalization of
ideas first suggested by \cite{white75} and subsequently developed
by others.  Here, the basic mechanism for attenuation/dispersion
is that when the sample is compressed the pore pressure in the
patch containing the stiffer fluid tries to equilibrate with that
in the more compressible patch by means of fluid flow from the one
region to the other. The effect is maximized if one of the fluids
(e.g. air) is much more compressible than the other (e.g. water).
In this Letter we use this theory of the Acoustics of
Patchy-Saturation (APS) to analyze measurements by \markcite{{\it
Cadoret et al.}} [1993, 1995, 1998] of wave velocities and
attenuation in partially water saturated limestones. Specifically,
as a function of fractional water-saturation, $S$, we deduce the
values of $V/A$, the ratio of the volume of the water patch to the
bounding area, and of the Poisson size, $l_f$, a different, but
equally well-defined, measure of patch size.  (In the special case
that the water patch is a d-dimensional hypersphere of radius $R$,
for example, one has $V/A =R/d$ and $l_f = R/\sqrt{d(d+2)}$.) It
is in this sense that one can ``hear" information about the size
and the shape of the patch; the wavelength of the slow
compressional wave is being used as the yardstick. As far as we
are aware, there is no other technique that could allow one to
make this kind of a deduction.

The APS theory makes several simplifying assumptions which
ultimately restrict the kinds of samples which can be analyzed by
it: (1) The Biot theory is presumed to be the only significant
mechanism for attenuation/dispersion.  This assumption rules out
the applicability of the APS theory to sandstones, for example,
which are known to be dominated by so-called microscopic squirt
mechanisms. (2) The measurement frequencies are so low that within
each patch the Biot theory is strictly in the low-frequency limit:
$\omega\ll\omega_B$, where
\begin{equation}
\omega_B=\frac{\eta\phi}{k\rho_f\alpha_\infty} \, ,
\end{equation}
in terms of the fluid viscosity, $\eta$, the porosity, $\phi$, the
permeability, $k$, the fluid density, $\rho_f$, and the tortuosity
of the pore space, $\alpha_\infty$.  For samples of high
permeability, this restricts the applicability of the theory to
rather low frequencies. (3) The frequencies are low enough that
the wavelengths are large compared to a characteristic patch size,
$L$: $\omega\ll\omega_x$ where
\begin{equation}
\omega_x=\frac{2\pi V_{\rm sh}}{L} \, . \label{w_x}
\end{equation}
$V_{\rm sh}$ is the velocity of the shear wave.  For samples with
large patch sizes, this also restricts the validity of the theory
to rather low frequencies. (4) Capillary effects, or rather the
change in capillary forces induced by the acoustic wave, are
neglected. Capillary effects clearly become more significant for
smaller pore sizes, i.e. for samples of lower permeability.
Moreover, if there is a very broad distribution of pore sizes,
capillary effects can lead to a very ramified, ``fractal",
structure for the patch geometry.  (5) The sample is
macroscopically homogeneous, except for the saturation patches.

Subject to the approximate validity of these assumptions, the APS
theory describes how the dynamic bulk modulus,
$\tilde{K}(\omega)$, crosses over from the Biot-Gassmann-Woods
result at low frequencies to the Biot-Gassmann-Hill result at
high:
\begin{equation}
\tilde{K}(\omega)=K_{\rm BGH}-\frac{K_{\rm BGH}-K_{\rm
BGW}}{1-\zeta+\zeta\sqrt{1-\imath\omega\tau/\zeta^2}} \,.\label{K}
\end{equation}
The low and the high frequency limits, $K_{\rm BGW}$ and $K_{\rm
BGH}$ respectively, depend upon the usual Biot parameters, as well
as the value of the saturation, but do {\it not} depend upon the
patch geometry.  We note for future use that if the gas phase is
taken to be infinitely compressible, then $K_{\rm BGW}(S) \equiv
K_b$ (the bulk modulus of the solid frame) for all values of saturation,
$S<1$.  The parameters $\zeta$
and $\tau$ {\it do} explicitly depend upon the two patch geometry
parameters $V/A$ and $l_f$, as well as the usual Biot parameters,
cf. Eqs.~(44) and (45) of \cite{johnson01}. It was demonstrated in
\cite{johnson01} that Eq. (\ref{K}) gives a very good
approximation for the dynamic modulus of patches that are slabs
(d=1), cylinders (d=2), spheres (d=3), or hollow spheres.  There
are no adjustable parameters in the theory.  In the following, we
assume that one knows all the Biot parameters for the system and
we investigate the effects of the partial saturation.  Our
procedure is to fit the relevant experimental data with Eq.
\ref{K}, thereby extracting values for $\zeta$ and $\tau$.  With
the values of $\zeta$ and $\tau$ one can easily solve for $V/A$
and $l_f$, for each value of the saturation.  In order to
demonstrate this, we analyze the experimental findings of
\markcite{{\it Cadoret et al.}} [1993, 1995, 1998].

We need data on samples which satisfy the aforementioned
assumptions and for which acoustic measurements were taken at
widely separated frequencies.  As far as we are aware the best
data for our purposes is that of Cadoret who performed two sets of
experiments on eight types of limestone (see \callout{Table~I}).
This suite of samples spans a wide range of permeabilities, even
though the porosities are all quite high. First, he measured the
velocity and attenuation of extensional and shear waves using the
resonant bar technique.  The samples were obtained by drilling out
(parallel to the rock bedding plane) a cylindrical rod with a
length of 110 cm and a diameter of 8 cm. The
lowest resonant harmonic frequencies are in the range 1-2 kHz, as
shown in Table~I.  Second, he performed ultrasonic
measurements of the compressional wave velocity at 50 and 100 kHz.
(He also reported measurements at 0.5 and 1.0 MHz, which we do not
use here.) In both sets of experiments, the samples were desaturated
by drying technique. By means of a careful analysis of these results,
as well as those taken on samples which were desaturated by a
depressurization technique, \markcite{{\it Cadoret et al.}} [1993,
1995, 1998] make a convincing case that the dominant mechanism for
dispersion and attenuation in the sample desaturated by drying
is the acoustically induced flow from one patch to the other.

\begin{figure}
\epsfig{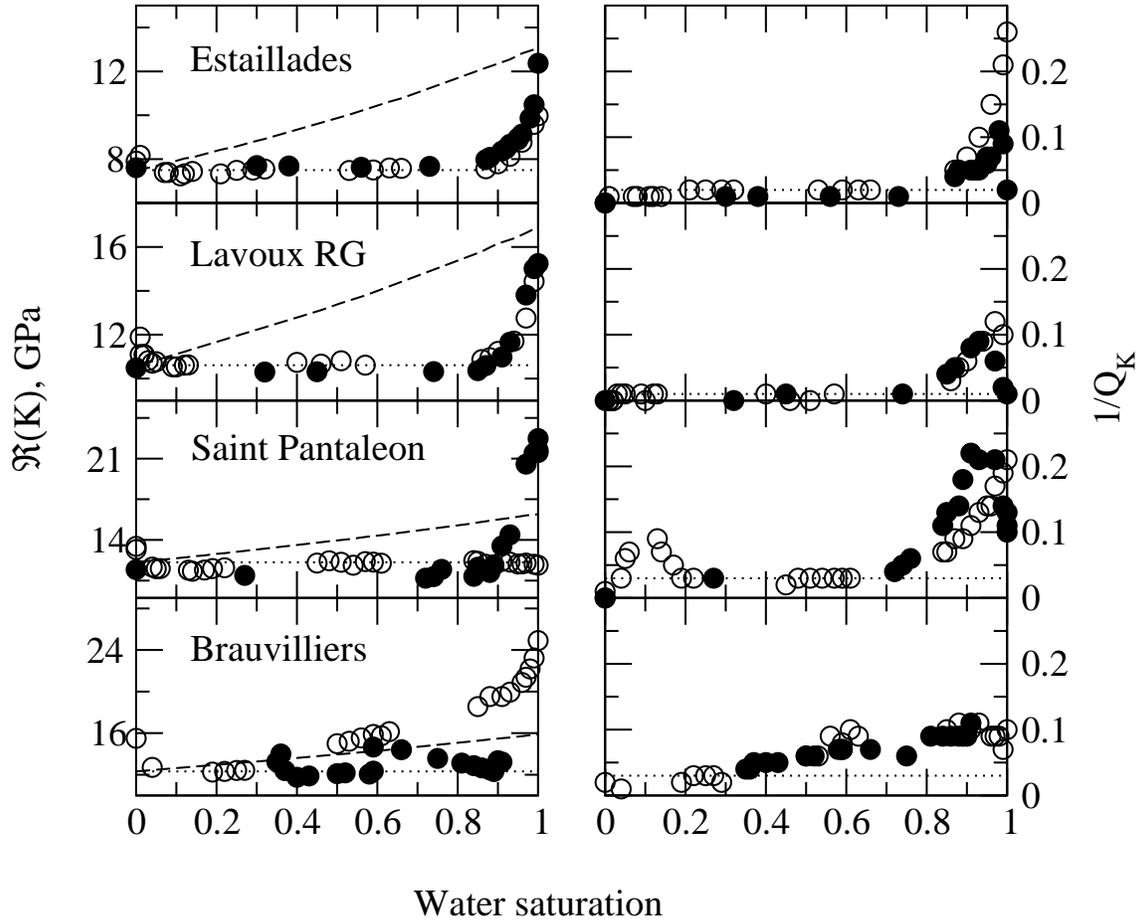}
\caption{Left plots show dispersion and right plots show
attenuation at sonic frequencies of patchy-saturated limestones.
Open and filled circles show experimental results which we extract
directly from the measurements of \markcite{{\it Cadoret}} [1993]
of the velocities and attenuation of extensional and shear waves.
Open (filled) circles correspond to the open (sealed) interface. On the left plots: the
dashed line is the high-frequency limit, $K_{\rm BGH}$, the
dotted line is the low-frequency limit, $K_{\rm BGW}$. On the right plots: the dotted line marks attenuation due to non-patchy-saturation mechanisms. All dashed and dotted lines in the figure correspond to the open interface case.} \label{F1}
\end{figure}

Our first step is to use the sonic data to deduce values for the
complex-valued bulk modulus $\tilde{K}(\omega)$ at the measurement
frequency. Then we use Eq. (\ref{K}) to find $V/A$ and $l_f$ as
described above. \callout{Fig.~\ref{F1}} shows
the bulk modulus dispersion and attenuation found from the
resonant bar data of Cadoret for four kinds of limestone.
Our convention is
\begin{equation}
\tilde{K}(\omega)=\Re(\tilde{K})(\omega)\left(1-\frac{\imath}{Q_K(\omega)}\right)
\, .
\end{equation}
We find $K_b$ from the
$\Re(K)$ data by finding the ``flat'' region of its saturation
dependence at $S<0.5$. It is marked by the dotted line. (The shear
modulus $N$ is found analogously from $\Re(N)$ data which we do
not show here.) The value of $K_b$ found
this way is different from the bulk modulus of the completely
dried frame (as one can see from the top plot) since a very small
amount of water leads to a noticeable softening of the sample.
Since the porosity, $\phi$, the bulk and shear frame moduli, $K_b$,
$N$ and the solid modulus $K_s$ ($\approx 69$ GPa for limestone)
are now known, it is straightforward to calculate $K_{\rm BGH}$ as a
function of saturation.  This is plotted as a dashed line.
The dotted line on the second plot of
Fig.~\ref{F1} shows the``flat'' region of the saturation
dependence of the attenuation $1/Q_K$ at $S<0.5$. The mechanism of
this attenuation is clearly different from and additional to the
patchy-saturation mechanism. In our analysis we subtract off this
contribution to $1/Q_K$ before solving for $V/A$ and $l_f$.

In the left plots of Fig.~\ref{F1}, the data lie between the high and low
frequency limits for the Estaillades and Lavoux RG limestones.
This is expected from the theory. On the other hand, the theory breaks
down in the case of the other two samples. We do the same analysis
for the rest of the limestones listed in
Table~I. We separate them into three categories, corresponding to
low, medium and high permeability. All of these limestones have
similar porosities and all have calcite mineralogy. We find that
the APS theory is mostly applicable to the rocks with medium
permeability (Lavoux RG, Estaillades and M\'enerbes). In the following,
we analyze data of Fig.~\ref{F1} for the Estaillades and Lavoux RG limestones
and then discuss why the theory does not work in the case of the high and low
permeability samples.

As discussed in \cite{johnson01}, the APS theory can apply if the
surface of the sample is either sealed or open as long as the gas
phase may be considered to be infinitely compressible.  Since the
experimental procedure for de-saturating the samples was different
in the two cases, the parameters $V/A$ and $l_f$ may also be
expected to be different in the two cases, for a given value of
$S$. In the sealed interface measurements, the boundary of the
fully water saturated rock was first jacketed and then the sample
was dried by making a few small open holes. This leads to a more
heterogeneous patch structure in the sample; while completely open
surface drying leads to air patches more or less contiguous
throughout the system, drying with only a few open holes makes air
saturate the sample preferentially near the holes, which are far
apart. The latter situation leads to larger patch sizes and, as a
result, a tighter limitation on the theory through Eq.
(\ref{w_x}).  That is, the relevant value of $L$ may be comparable
to the length of the rod, rather than to the radius.  We think the
case of a closed interface lies beyond the applicability of the
APS theory as $\omega$ exceeds $\omega_x$ and we have not pursued
it any further.

In Table~I we show the Biot crossover frequency $\omega_B$ as well
as $\omega_x$. The latter is evaluated by setting $L=2R$ in Eq.
(\ref{w_x}). We expect this to be an overestimate of the
patch sizes in the case of open interface, but in the case of
closed interface with high saturation values, this can be an
underestimate, as discussed above. (See \markcite{{\it Cadoret et
al.}} [1993, 1995, 1998] for images of the computerized tomography
scans of fluid distribution within the rock.)

\begin{figure}
\epsfig{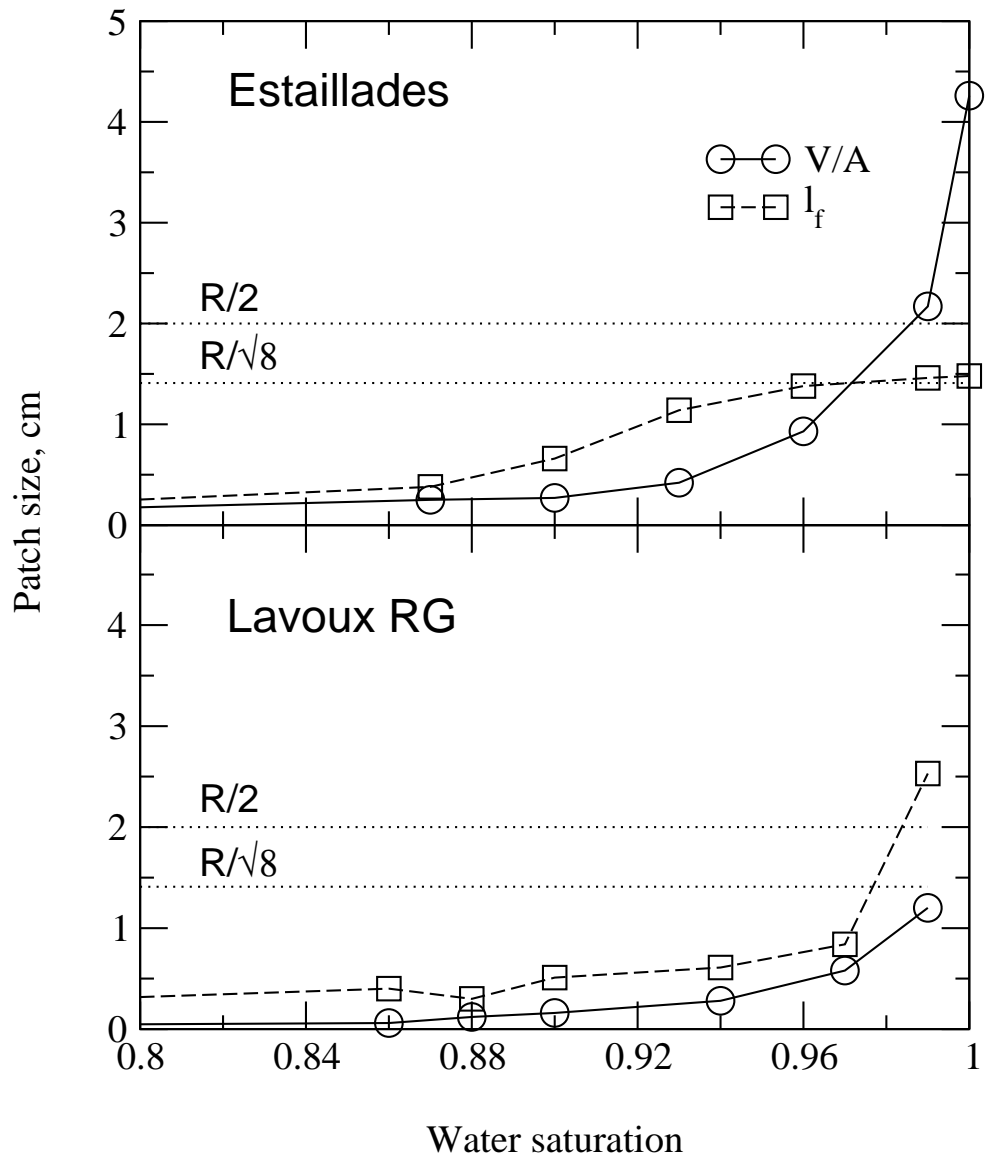}
\caption{Characteristic water patch sizes calculated from
the dispersion and attenuation data in Fig.~\ref{F1}
using the APS theory.} \label{F2}
\end{figure}

\callout{Fig.~\ref{F2}} shows characteristic patch sizes extracted
from Cadoret's measurements on the open interface. The dotted
lines on the plot indicate the value $R/2$ ($R/\sqrt{8}$) of the
sample, which is the expected value of $V/A$ ($l_f$) as
$S\rightarrow 1$. We do not know why some of our deduced values of
these parameters exceed these limits at high saturation, $S>0.99$;
it may be due to the neglected capillary effects or
heterogeneities in the value of the permeability. Notwithstanding,
the fact that our analysis comes close to the expected values and
gives estimated patch sizes which decrease as the saturation
decreases is, we feel, significant substantiation of our approach
in this Letter.

\begin{figure}
\epsfig{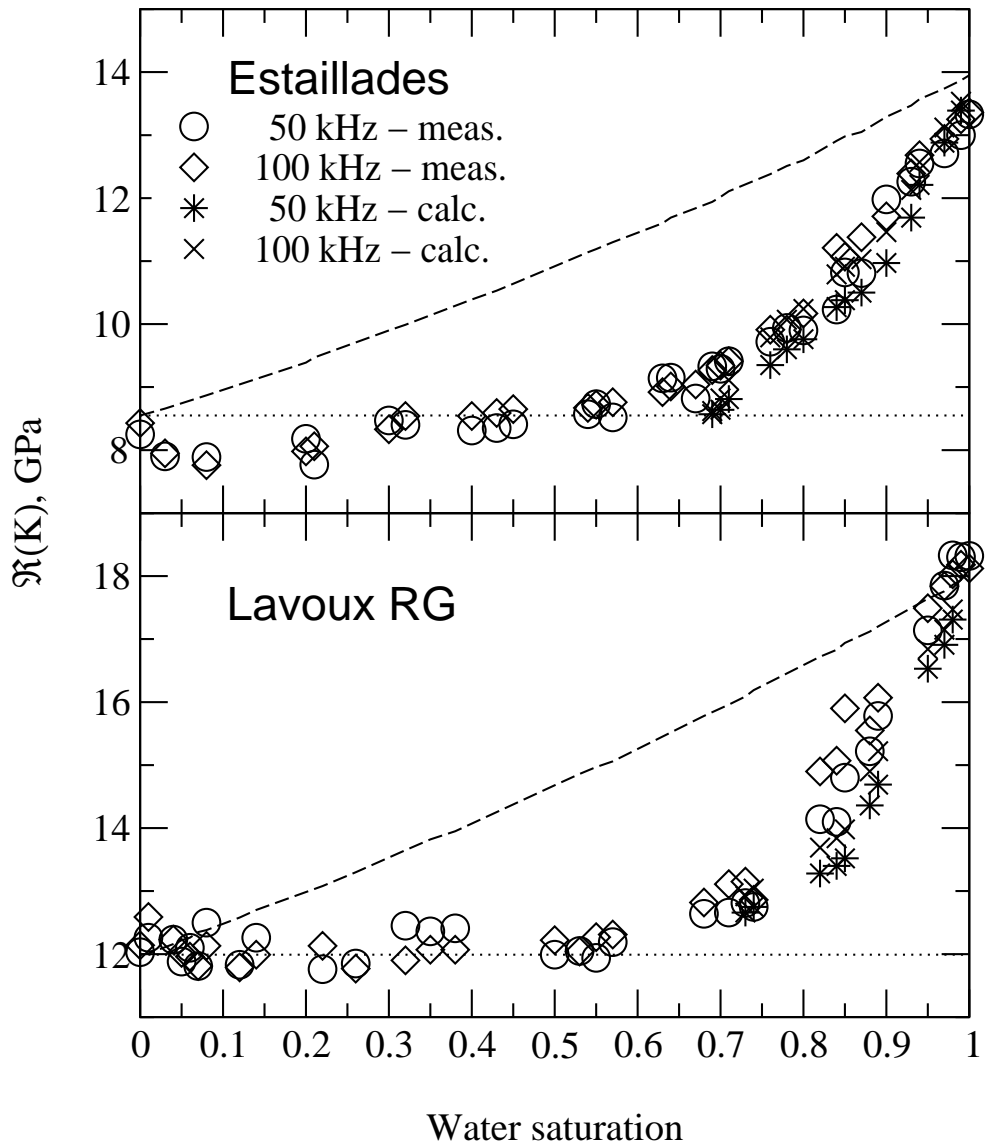}
\caption{Circles and diamonds show ultrasonic dispersion
extracted from Cadoret's measurements of compressional wave
velocities; stars and crosses show corresponding calculated
results using parameters from Fig.~\ref{F2}.} \label{F3}
\end{figure}

It is clear from this figure that the deduced values of the patch
sizes are quite plausible.  However, at each saturation, two
experimentally determined numbers have been used to deduce the two
unknowns, $V/A$ and $l_f$.  As a check, we use our deduced values
of $V/A$ and $l_f$ as functions of saturation to calculate
$\Re(\tilde{K})(\omega)$ at ultrasonic frequencies, 50 and 100
kHz, and compare it against the measurements of the compressional
wave velocities by \markcite{{\it Cadoret et al.}} [1993, 1995,
1998]. The attenuation was not measured. \callout{Fig.~\ref{F3}}
shows our findings. There is a reasonable agreement, but one
should be cautious since the ultrasonic frequency regime lies
barely within the scope of the APS theory, as can be seen from
Table~I: $\omega_B$ lies within the same frequency decade as both
50 kHz and 100 kHz where the quasistatic approximation of the Biot
theory starts to break down, $\omega_x$ also lies in the same
decade, but, as noted above, this is an overestimate for the open
interface case and the APS theory should still give a reasonably
good approximation even when $\omega\sim\omega_x/2$. Also, by
comparing the ``flat'' regions (marked by dotted lines) in the top
and bottom plots, one can observe that the frame bulk modulus,
$K_b$ itself scales with frequency. This effect lies outside the
scope of the APS theory by invalidating assumption (1).
Nevertheless, we report a good agreement between measured and
calculated values of $\Re(\tilde{K})$ as seen in Fig.~\ref{F3}.

The APS theory appears not to apply to the samples with either
very high or very low permeability.  First, the rocks with high
permeability (Espeil and Saint Pantaleon) have a low Biot
frequency (Table~I) and even the sonic measurements are separated
from $\omega_B$ by only a decade in frequency. The inconsistency
of the APS theory in this regime can be seen from Fig.~\ref{F1};
the real part of the bulk modulus exceeds $K_{\rm BGH}$ at high
saturation in the case of closed interface, but within the APS
theory, $\Re(\tilde{K})$ is always less than $K_{\rm BGH}$. Also,
in the case of the open interface, $\Re(\tilde{K})$ has very
little variation with saturation, which implies the low frequency
regime, but the significant attenuation for $S>0.8$ contradicts
this.  As was shown in \cite{johnson01} for frequencies comparable
to $\omega_B$, the attenuation predicted by the full Biot theory
can be much larger than that predicted by the APS theory.  The
theory should not work here, and it doesn't.

Second, the rocks with low permeability (Lavoux RF, Bretigny and
Brauvilliers) also have experimental values of the bulk modulus
inconsistent with the APS theory; this can be observed in
Fig.~\ref{F1} where we see that the real part of $\tilde{K}$
exceeds $K_{\rm BGH}$.  It is true that the neglected capillary
forces may act to pin the water-air interface but, within the
context of the Biot theory, the modulus cannot exceed $K_{\rm
BGH}$.  Rather, the data are suggestive of a non-Biot mechanism,
such as microscopic squirt, though we have no specific evidence
for this.

In conclusion, we have shown how the APS theory \cite{johnson01}
of frequency-dependent acoustics in patchy-saturated media can be
used to extract information about characteristic fluid patch sizes
in partially saturated rocks.  We demonstrated this by analyzing
resonant bar measurements by \markcite{{\it Cadoret et al.}}
[1993, 1995, 1998] of the acoustic properties of eight types of
limestone.  For limestones with intermediate values of
permeability, the APS theory describes the dominant mechanism of
dispersion and attenuation due to patchy-saturation of the sample.
Limestones with high values of permeability lie outside the scope
of the quasistatic Biot theory and, therefore, invalidate the
assumptions of the APS theory, while in the limestones with low
permeability, there is another significant mechanism of dispersion
and attenuation.   Strictly speaking, the ultrasonic measurements
of Cadoret lie outside the range of validity of the quasistatic
Biot assumption for all eight types of limestone.  Nonetheless, we
conclude that one can still deduce patch size and shape
characteristics of the saturation pattern from the acoustic
measurements, when the assumptions inherent to the APS theory are
maintained in the experiments.

This work was supported in part by NSF grant DMR 99-81283.

\end{article}

\begin{planotable}{llcccccc}
\tablehead{$k$ regime & Limestone & Porosity, \% & $k_H$\tablenotemark{a},
mD\tablenotemark{b} & $k_V$\tablenotemark{a}, mD\tablenotemark{b} &
$\omega/2\pi$\tablenotemark{c,d} & $\omega_B/2\pi$\tablenotemark{c} &
$\omega_x/2\pi$\tablenotemark{c}}
\tablecaption{Characteristics of the limestones used in the
measurements of \markcite{{\it Cadoret}} [1993].}
\tablenotetext{a}{$k_H$ is horizontal and $k_V$ is vertical permeability.}
\tablenotetext{b}{One mD = $10^{-11}cm^2$.}
\tablenotetext{c}{Frequencies are in kHz.}
\tablenotetext{d}{Resonant frequency $\omega$ corresponds to the
fully-saturated sample.}
\startdata
low & Lavoux RF & 24 & 7.5 & 8.9 & 1.3 & 2400 & 23 \\
& Bretigny & 18 & 15.0 & 2.0 & 1.7 & 800 & 30 \\
& Brauvilliers & 41 & 15.4 & 2.1 & 1.2 & 2800 & 22 \\
medium & Lavoux RG & 24 & 16 & 44 & 1.4 & 400 & 25 \\
& Estaillades & 30 & 255 & 269 & 1.2 & 100 & 22 \\
& M\'enerbes & 34 & 320 & 288 & 1.1 & 99 & 21 \\
high & Espeil & 28 & 1853 & 1211 & 1.1 & 13 & 20 \\
& Saint Pantaleon & 35 & 4067 & 2141 & 1.1 & 8.2 & 21
\end{planotable}

\begin{thebibliography}{}
\bibitem[{\it Biot,} 1954]{biot54}
\reference Biot, M. A.,
Theory of propagation of elastic waves in a fluid-saturated porous solid,
{\it J. Acoust. Soc. Am.,} {\it 28,} 168-191, 1954.
\bibitem[{\it Cadoret,} 1993]{cadoret93}
\reference
Cadoret, T., Effet de la saturation eau/gaz sur les propri\'{e}t\'{e}s
acoustiques des roches; \'{e}tude aux fr\'{e}quences sonores et
ultrasonores, Ph.D. thesis, 250 pp., L'Universit\'{e} de Paris VII,
December 1993.
\bibitem[{\it Cadoret et al.,} 1995]{cadoret95}
\reference Cadoret, T., D. Marion, B. Zinszner,
Influence of frequency and fluid distribution on elastic wave velocities in
partially saturated limestones,
\jgr {\it 100,} 9789-9803, 1995.
\bibitem[{\it Cadoret et al.,} 1998]{cadoret98}
\reference Cadoret, T., G. Mavko, B. Zinszner,
Fluid distribution effect on sonic attenuation in partially saturated
limestones,
{\it Geophysics,} {\it 63,} 154-160, 1998.
\bibitem[{\it Johnson,} 1986]{johnson86}
\reference Johnson, D. L.,
Recent developments in the acoustic properties of porous media, in {\it
Frontiers in Physical Acoustics XCIII,} edited by D. Sette, pp. 255-290,
North Holland Elsevier, New York, 1986.
\bibitem[{\it Johnson,} 1994]{johnson94}
\reference Johnson, D. L., Plona, T. J., and Kojima, H., Probing
Porous Media with First and Second Sound II. Acoustic Properties
of Water-Saturated Porous Media, {\it J. Appl. Phys., 76},
115-125, 1994.
\bibitem[{\it Johnson,} 2001]{johnson01}
\reference Johnson, D. L., Theory of frequency dependent acoustics
in patchy-saturated porous media, {\it J. Acoust. Soc. Am.,} (to
appear) 2001.
\bibitem[{\it White,} 1975]{white75}
\reference White, J. E., Computed Seismic Speeds and Attenuation
in Rocks with Partial Gas Saturation, {\it Geophysics,} {\it 40},
224-232, 1975.


\end{thebibliography}
\end{document}